\documentclass[twocolumn,twoside]{tdc2}
\usepackage[dvips]{graphicx}
\usepackage{here}
\usepackage{tdc_macro}
\begin{document}
\twocolumn 

\begin{mtitle}
KNIFE, Kashima Nobeyama InterFErometer
\end{mtitle}

\begin{mauthor}
Makoto Miyoshi ({\it makoto.miyoshi@nao.ac.jp})
\vspace{\baselineskip}
{\it National Astronomical Observatory, Japan, 2-21-1, Osawa, Mitaka, Tokyo, Japan, 181-8588}
\end{mauthor}

\setcounter{section}{0}
\setcounter{subsection}{0}
\setcounter{figure}{0}
\setcounter{table}{0}
\setcounter{footnote}{0}

\begin{figure*}[ht]
\begin{center} 
\includegraphics[width=0.9\textwidth,scale=1,angle=0]{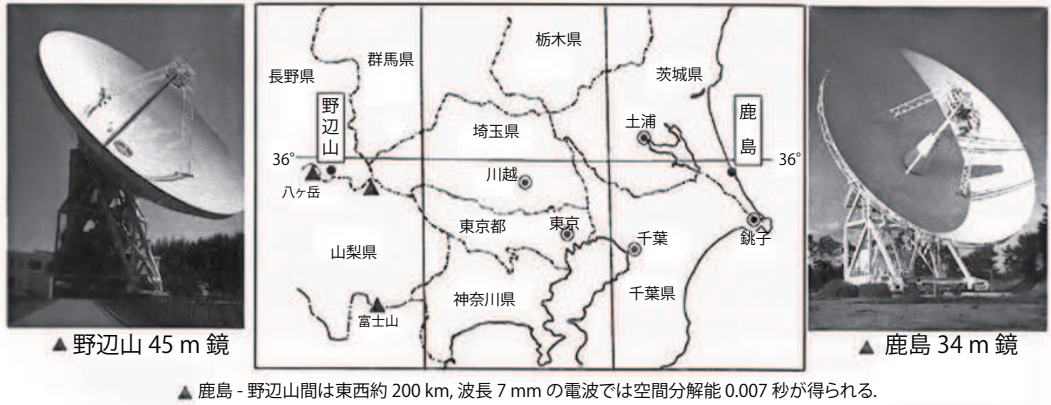}
\end{center} 
\caption{KNIFE baseline: The baseline length is 197.66~km. The north-south component is only 240~m.
The fringe spacing is about 7~mas at 43~GHz observations, suitable for SiO maser observations.}
\label{Fig-1}
\end{figure*}

\hspace{-1em}{\it Abstract}:
By connecting two antennas, Kashima 34~m and Nobeyama 45~m, an east-west baseline of 200~km is formed. At that time, because Nobeyama 45~m had the world's number one sensitivity in the 43~GHz band, and also Kashima 34~m was the world's third-largest one, the Kashima-Nobeyama baseline provided the highest sensitivity at 43~GHz VLBI (Figure~\ref{Fig-1}). 
The construction of the Kashima 34~m antenna began in 1988,  also almost at the same time, a domestic project of mm-VLBI  (KNIFE, Kashima Nobeyama INterFrermeter) started. Nobeyama Radio Observatory provided the first cooled-HEMT 43~GHz receiver in the world to the Kashima 34~m. In October 1989, the first fringe at 43~GHz was detected. We here review the achievements of the KNIFE at that time.

\section{KNIFE project}
\begin{figure}[ht]
\begin{center} 
\includegraphics[width=0.45\textwidth,scale=1,angle=0]{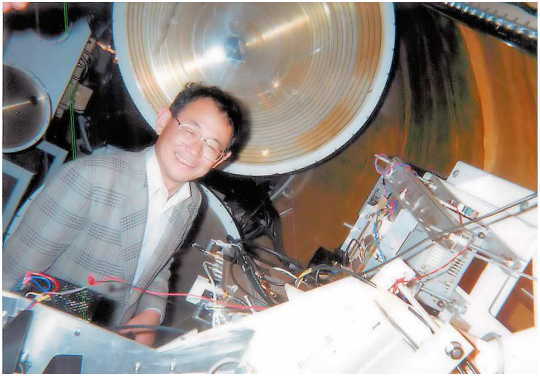}
\end{center} 
\caption{Prof. Masaki Morimoto, who promoted the KNIFE project.}
\label{Fig-2}
\end{figure}

\begin{figure}[ht]
\begin{center} 
\includegraphics[width=0.45\textwidth,scale=1,angle=0]{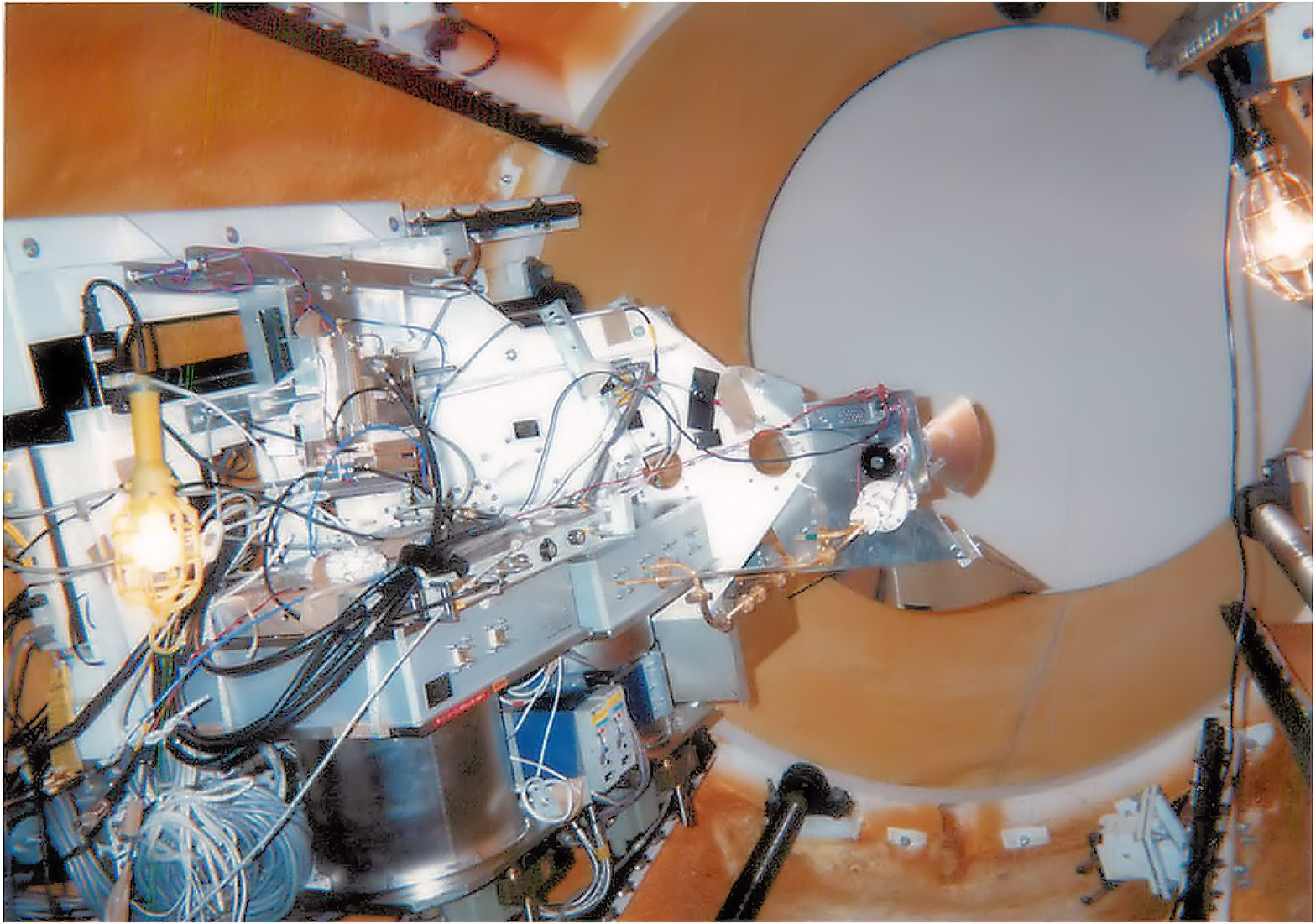}
\end{center} 
\caption{The cooled HEMT Receiver installed to 34~m telescope. This photo was taken at the time when the first 43~GHz fringe was detected in Oct. 1989. }
\label{Fig-3}
\end{figure}

In 1987, the Radio Research Laboratory (RRL, NICT at present) decided to construct the Kashima 34~m antenna as the main station of the Western Pacific interferometer.
At that time, the Nobeyama Radio Observatory, a branch of the National Astronomical Observatory of Japan (NAOJ) participated in the global millimeter-wave VLBI using the Nobeyama 45~m and had just started VLBI observations. A set of  Mark-3 recorder was brought to the Usuda station from Nobeyama, and the first space VLBI experiment with the TDRS satellite was conducted using Usuda 64~m, and the fringe detection was successful. However, in Japan, independent astronomical VLBI observational research has not yet been possible.
In response to the news of the construction of the Kashima 34~m antenna, Prof. Morimoto (Figure~\ref{Fig-2}) noticed that the Kashima 34~m antenna has a  surface accuracy of $170~\mu \rm as$ and that it is very effective for millimeter-wave VLBI observation. 
Prof. Morimoto proposed to RRL to conduct the millimeter-wave VLBI researches in collaboration with NAOJ, and the joint research began. Using the 43~GHz cooling receiver dewar owned by RRL, NAOJ decided to manufacture the world's first 43~GHz cooled HEMT receiver (Figure~\ref{Fig-3}) and the joint research started in 1989.
The KNIFE experiment started at the same time as the start-up and test of the 34~m antenna. Though the first KNIFE experiment on June 14-15, 1989 failed, at the second KNIFE experiment on October 20-22, 1989, the first 43~GHz fringes were successfully detected from the observations of the strong SiO masers in  Orion~KL and VY~CMa.


\section{SiO maser observations}
The SiO maser emissions originate at circumstellar envelopes of late-type stars, but they were mysterious phenomena at that time.
VLBI observations of SiO masers had been carried out, but the angular size of SiO maser was quite large, and then resolve-outs were occurred at long baselines, so the structures of SiO masers were not well understood. The KNIFE baseline is 200~km. At 43~GHz, the minimum fringe spacing is about 7 mas, so it is just suitable for observing SiO masers. Prof. Morimoto urged the install of the 43~GHz receiver on the Kashima 34~m antenna as soon as possible because fruitful scientific results were expected.

\begin{figure*}[ht]
\begin{center} 
\includegraphics[width=0.9\textwidth,scale=1,angle=0]{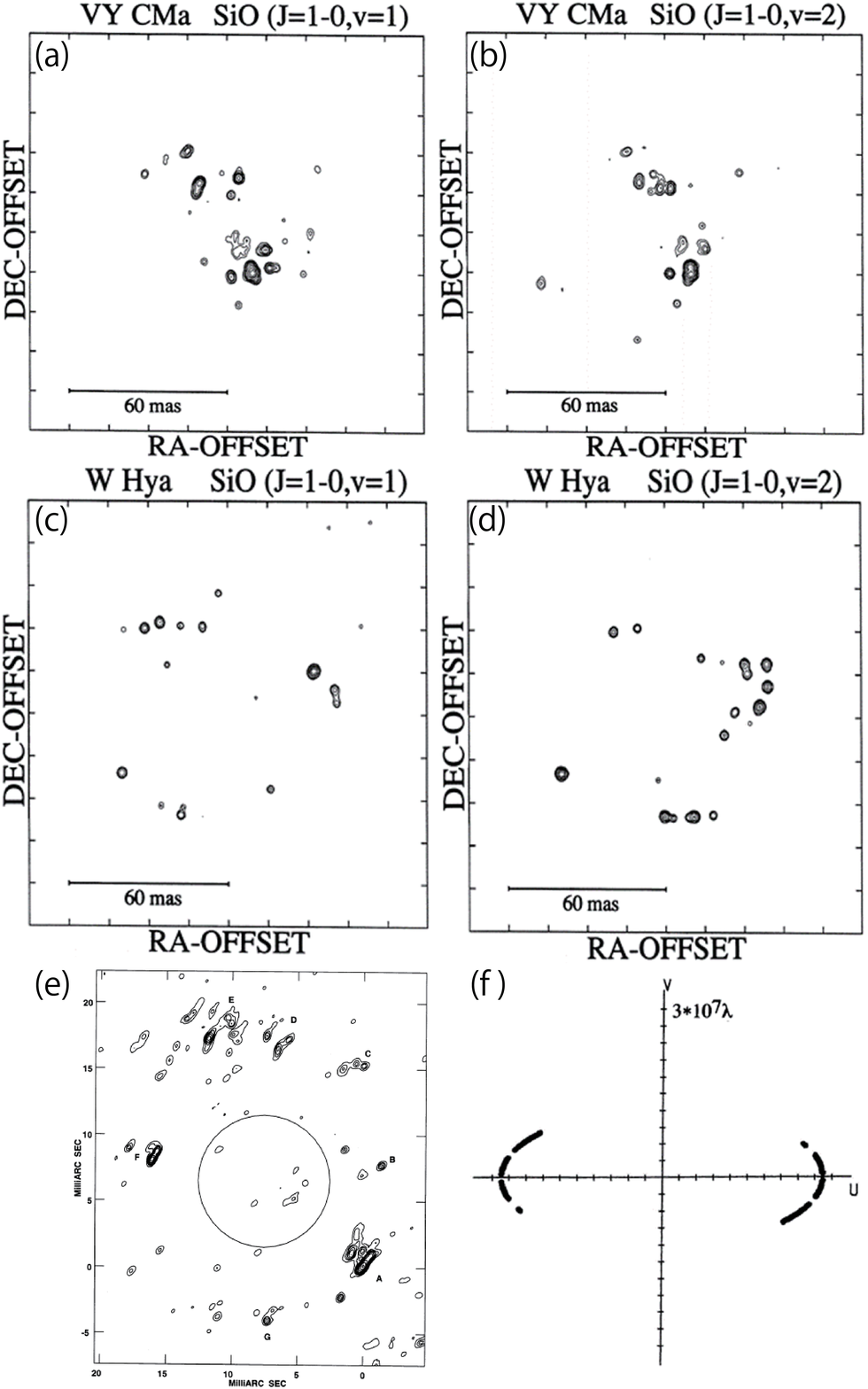}
\end{center} 
\caption{
SiO masers observed with KNIFE.
(a) VY CMa J=1-0, v=1,
(b) VY CMa J=1-0, v=2,
(c) W Hya J=1-0, v=1,
(d) W Hya J=1-0, v=2,
(e) U Her observed with VLBA \cite{D1994}, (f) u-v coverage of KNIFE baseline for them.
}
\label{Fig-4}
\end{figure*}

The Figure~\ref{Fig-4} shows the observational result of SiO maser by KNIFE. 
The star itself cannot be detected, but the SiO masers are distributed around the star. 
The distributions of SiO masers with $v = 1$ and $v = 2$, though their excitation temperatures differ about 1800~K, are almost the same as each other. This means that the maser is inferred to be due to collisional excitation \cite{Miyoshi1994}. 
At that time, the VLBA in USA was in the process of being constructed, but SiO maser observations were performed as a test.
Panel (e) in the Figure ~\ref{Fig-4} shows the VLBA result (SiO masers in U~Her~\cite{D1994}).
If compared to the result of SiO masers by the VLBA four stations,
It can be understood that the KNIFE performance was considerably good even though it was a single baseline.

\section{Observations of high velocity water mega maser in NGC~4258}
\begin{figure*}[ht]
\begin{center} 
\includegraphics[width=0.9\textwidth,scale=1,angle=0]{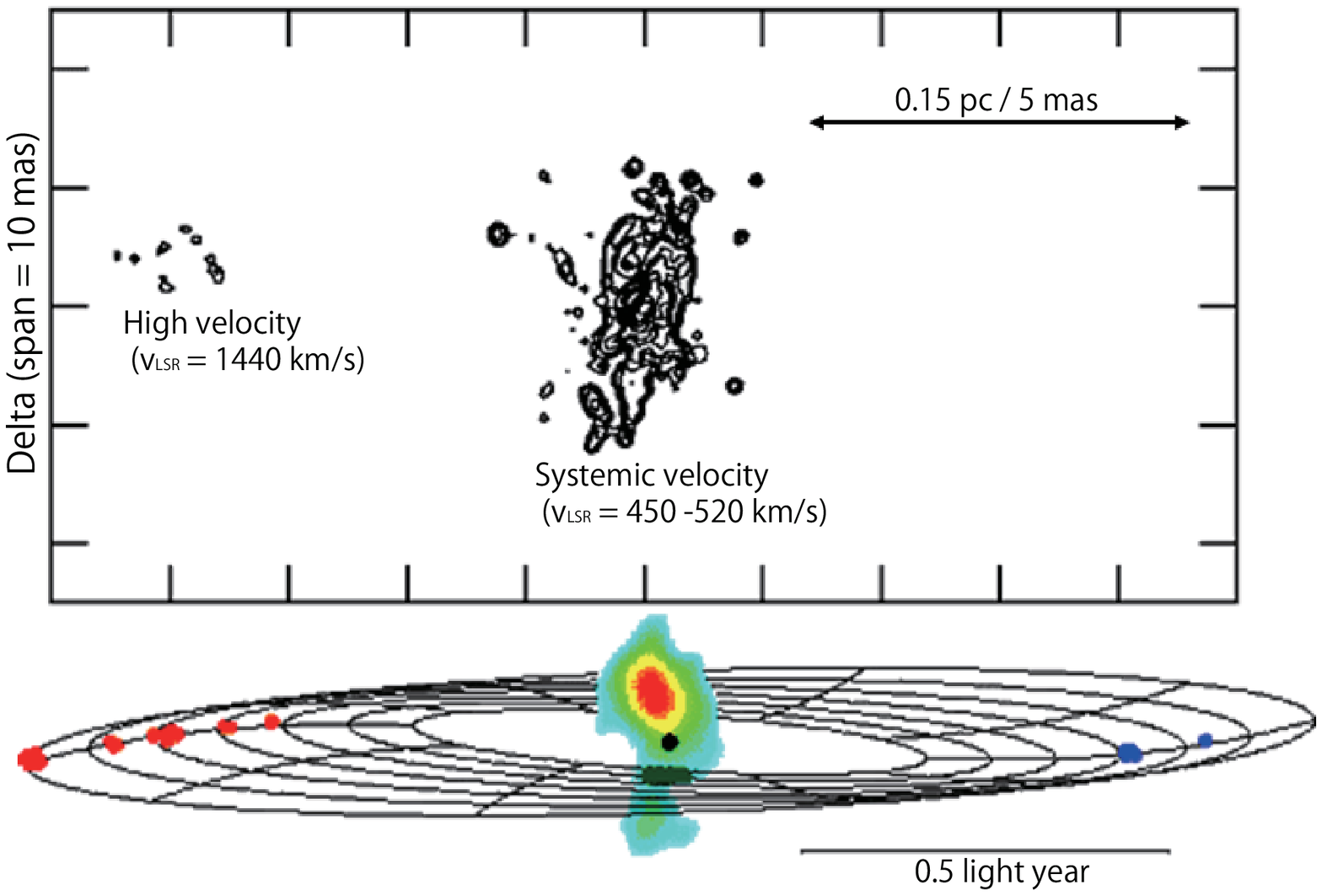}
\end{center} 
\caption{
The oldest map of high velocity water mases in NGC~4258 observed with KNIFE in 1992 June (top panel).
The image was produced in 1997. This map demonstrates that the KNIFE had the performance to clarify the structure of water mega masers in  NGC~4258.
The bottom panel shows the image with VLBA observations~\cite{H1999}.
}
\label{Fig-5}
\end{figure*}

In 1992, Nakai et al discovered high-velocity components in the water mega maser of the galaxy NGC~4258 by using the Nobeyama 45~m in of Japan~\cite{Nakai1993}.
Relative to the known water maser components, they show velocity shifts of $\pm1000~\rm km/s$. 
There are two groups, one shows blue shifts of about 1000~ km/s while the other shows red shifts of about 1000~km/s.
 Hearing the news, Prof. Morimoto said, "If the masers move with velocities about 1000~km/s, we can easily detect the proper motions of the maser components. Observe with Kashima 34~m immediately! Observe twice so as not to fail!" 
What is the origin of the high-velocity mega masers of NGC~4258? 
In order to identify the origin, observations with a high spatial resolution are required. Prof. Morimoto suggested that we should investigate with KNIFE as soon as possible.
The KNIFE observations were conducted in early June 1993, less than a month after the discovery. However, due to calculation errors of the observational sky frequencies, correlations by new NAOCO correlator in NAOJ~\cite{Shibata} that has not even been bug-fixed, and the use of imaging software made tentatively, it was not possible to find the maser distributions exactly. However, in August 1993, it is found that the high-velocity components are also located at the center of the galaxy, within 50~mas of the known main components.
In 1995, VLBA observations revealed that the water mega maser in NGC~4258 was from a molecular gas disk with Keplerian  motion around a supermassive black hole. 
From the maser velocity and structure, the mass of the black hole is estimated to be $3.6\times 10^7\rm~M_{\odot}$~\cite{Miyoshi1995}.
The Figure~\ref{Fig-5} shows the mega maser distribution obtained from reanalyzing the KNIFE data and that from the result of VLBA. 
As you can see from the Figure, it is possible to capture the structure of the water mega maser in NGC~4258 even with the single 200~km baseline of KNIFE. The KNIFE observation of NGC~4258 also shows the excellent performance of the KNIFE.

\section{The role played by the KNIFE project}
The results are not only those mentioned above. Pioneering research was conducted on higher-frequency geodetic VLBI using KNIFE~\cite{Matumoto1994}. 
In addition, survey observations of water and SiO masers were carried out using the single Kashima 34~m antenna and KNIFE ~\cite{Takaba1994}, \cite{Imai2001}.
The KNIFE astronomical observations using the single-baseline between Kashima 34~m and Nobeyama 45~m were carried out until 1992. In 1993 it developed into J-Net observations with 4 stations including Mizusawa 10~m antenna and Kagoshima 6~m antenna; Monitoring VLBI observations of the burst phenomena of water masers in Orion KL were performed before its beginning by chance
\cite{OModaka}. 
KNIFE paved the way for the subsequent development of astronomical VLBI researches in Japan.

\end{document}